\documentclass[superscriptaddress, nofootinbib, 11pt]{revtex4} 
\usepackage{amsmath}
\usepackage{amsfonts,amssymb}
\usepackage[T1]{fontenc} %permite uso direto de caracteres com acento, i.e "ção" ao invés de \c{c}\~{a}o
\usepackage[latin1]{inputenc}

\newcommand{\roundbraket}[1]{ \left( {#1} \right) }

\newcommand{\req}[1]{eq.~(\ref{#1})}
\newcommand{\beq}{\begin{equation}} 
\newcommand{\eeq}{\end{equation}}

\newcommand{\nn}{\nonumber}

\newcommand{\eps}{\varepsilon}
\newcommand{\TVnorm}[1]{\left\| {#1} \right\|_{TV}}

\newtheorem{lemma}{Lemma}

\begin{document}
\title{Comment on the paper ``Random Quantum Circuits are Approximate 2-designs'' }
\date{\today}
\author{Igor Tuche Diniz}
\affiliation{Instituto de Física, Universidade Federal Fluminense, Niterói, Brazil}
\affiliation{Institut Néel-CNRS, Grenoble, France}
\author{Daniel Jonathan}
\affiliation{Instituto de Física, Universidade Federal Fluminense, Niterói, Brazil}
\email{jonathan@if.uff.br} 
\email{igor.diniz@grenoble.cnrs.fr}

\begin{abstract}

In [A.W. Harrow and R.A. Low, Commun. Math. Phys. 291, 257-302 (2009)], it was shown that a quantum circuit composed of random 2-qubit gates converges to an approximate quantum 2-design in polynomial time. We point out and correct a flaw in one of the paper's main arguments. Our alternative argument highlights the role played by transpositions induced by the random gates in achieving convergence.
\end{abstract}

\maketitle

\section{Introduction}

\emph{Quantum $k$-designs} \cite{Emerson07} are statistical ensembles over the sets of states or operators of a quantum system that faithfully reproduce the $k^{th}$ moments of the respective uniform distributions. These pseudo-random ensembles are of interest since they can often be efficiently simulated in a physical system. In other words, while physically generating random states or operators of an $n$-qubit quantum system requires resources that grow exponentially in $n$, pseudorandom objects may require only polynomial resources \cite{EWSLC03}. They are thus a practical tool for a wide variety of communication and computation tasks that make use of random quantum objects (e.g., \cite{Private,SEQPT,ObliSuperdense}).

In ref. \cite{HL08a}, Harrow and Low (HL) have provided an example of an \emph{efficient} construction of a quantum $2$-design for operators of an $n$-qubit system, i.e., one that can be physically implemented using resources that scale
polynomially with $n$. Unlike previous constructions with this property \cite{Divincenzo01,GAE07,Dankert09}, their scheme appears to be efficient also for higher values of $k$ \cite{BV09}. The construction is based on a random quantum circuit model \cite{EWSLC03}: at each step of the circuit, a pair of qubits is chosen at random, and a 2-qubit gate is applied to them, drawn from some ensemble $\mu$ over the set of all such gates. The pseudorandom $n$-qubit operators that result from this procedure have second moments whose evolution can be reduced to a classical Markov chain \cite{ODP,ODPa}. In particular, the (approximate) convergence of this chain to its stationary state is sufficient to ensure the convergence of the pseudorandom operator ensemble to an approximate quantum $2$-design \cite{HL08a}. 

In this note we wish to point out and correct a flaw in a significant step of this
analysis, on which the main results of ref. \cite{HL08a} directly depend. Specifically,
the proof of Corollary 5.1 (p. 284), a statement concerning the number of steps required for the convergence of the Markov chain, is incorrect. We give an alternative argument showing that the statement itself is indeed valid. Our proof highlights the role played by transpositions induced by the random gates in achieving convergence.

We assume that the reader is familiar with ref.
\cite{HL08a}. In section \ref{sec:summary} we summarize some of its results,
explaining where they are affected by the flawed step. In section \ref{sec:flaw} we
explain the flaw itself, giving an explicit counterexample. In section \ref{sec:prelims}
we give the general idea of our argument, and develop some preliminary results using
standard tools from Markov chain theory and group representations. Section
\ref{sec:proof} contains our main result, with several details left to the Appendix.

\section{Summary of results in \cite{HL08a} }\label{sec:summary}

Following a strategy introduced in \cite{ODP,ODPa}, the first part of ref. \cite{HL08a} establishes a map from the evolution of second
moments of a random quantum circuit to a classical Markov chain $P$ with state space
$\Omega_P =~\{0,1,2,3\}^n$. When the ensemble $\mu$ is chosen to be the
uniform (Haar) distribution over $U(4)$, $P$ turns out to have a particularly simple
form, described by the following algorithm: given a position $\vec{p} = (p_1, \ldots p_n)
\in \Omega_P $, choose a new position $\vec{p}'$ as follows: \beq \label{eq:P}
\begin{split}
   &-\text{choose randomly and uniformly a pair of indices $1\leq i \neq j \leq n$}. \\
   &- \text{if $p_i = p_j = 0$, do nothing }\\
   &-  \text{if $(p_i, p_j) \neq (0,0)$, replace the pair with any element of $\{0,1,2,3\}^2 \backslash {(0,0)}$, } \\
    & \text{   choosing uniformly from the 15 possibilities}.
\end{split}
\eeq
The corresponding Markov matrix $P(\vec{p}, \vec{p'})$  has the form $P = \frac{1}{n(n-1)}\sum_{i \neq j} P_{ij}$, where $P_{ij}$ affects only the $i,j$ coordinates of $\vec{p}$. Apart from an isolated stationary state $\vec{0} = (0 \ldots 0)$, this Markov chain is ergodic, with stationary state given by the uniform distribution
\beq \label{eq:Pstat}
\pi (\vec{p})  = (4^n -1 )^{-1}, \forall \vec{p} \in  \Omega_P \backslash \{\vec{0}\}.
\eeq
The key technical problem is then to analyze the convergence time of $P$, measured for example by its mixing time in the trace norm:
\beq\label{eq:tmix}
  t_{mixP}(\eps) : = \max_{\vec{p} \in \Omega_P \backslash \vec{0} } \left[\min \left\{t \left| \; \TVnorm{P^t(\vec{p}, \cdot)  - \pi}< \eps \right. \right\}\right].
\eeq

HL's approach is to concentrate first on the much smaller Markov chain $Z$ which tracks the number of nonzero coordinates (i.e., the Hamming weight $H(\vec{p}) = \left| \left\{ i \left| p_i \neq 0\right. \right\}\right| $ ) of states evolving under the $P$ chain. This `zero chain'  is ergodic on the state space $\Omega_Z= \{1, \ldots, n\} $, with stationary state
\beq \label{eq:statZ}
\zeta_{\pi} (H) = \frac{\binom{n}{H} 3^H}{4^n -1};\; \quad H \in  \Omega_Z
\eeq
Its only nonvanishing transition probabilities  are (eq. 5.2 in \cite{HL08a}):
\begin{equation}
\begin{split}
  &Z(H,H +1) =  \frac{3}{5} \; H  \; (n-H) \roundbraket{ n \atop 2 }^{-1}  \;,  \\
  &Z(H,H-1) =  \frac{1}{5} \; H  \; (H-1) \roundbraket{ n \atop 2 }^{-1} \;,  \\
  &Z(H,H) = 1 -  Z(H,H-1)  -  Z(H,H+1) = 1- \frac{2 H (3n -2H -1)}{5 n(n-1)} \;.
\label{eq:zetaDef}
\end{split}
\end{equation}
Determining a tight upper bound on the mixing time $t_{mixZ} (\eps)$ of this chain turns out to be quite tricky. The main difficulty is dealing with states with small values of $H$, which by \req{eq:zetaDef} only have probability $O(1/n)$ of evolving. Nevertheless, after a laborious calculation, HL are able to show in Theorem 5.1 that $t_{mixZ} (\eps)  = \Theta(n \log (n/\eps) )$.

The next step in the analysis is the one that concerns us in this Comment.  In Corollary 5.1, Harrow and Low state that, once the $Z$ chain has approximately mixed, then $O(n \ln( n/\eps) )$ further steps suffice to ensure the convergence of the $P$ chain as a whole, so that \smallskip

 \textbf{Corollary 5.1 \cite{HL08a}:} The full (P) chain mixes in time $t_{mixP} (\eps)  = \Theta \left(n \log\left(\frac{n}{\eps}\right)\right)$.
\smallskip

It is important to emphasize that, despite its moniker, this result is in fact an independent theorem that does not follow automatically from other results in \cite{HL08a}. It is also a vital step in the main argument of the paper, as it implies immediately (see eq. 5.7 and Theorem 4.1) that the spectral gap $\Delta$ of the $P$ chain is of order $\Theta (1/n)$. This fact is, in turn, necessary for the main conclusions of the paper, viz. Theorem 2.2 giving the polynomial bound for the convergence of a random quantum circuit to a 2-design. 

Unfortunately, as we now show, 
the demonstration of Corollary 5.1 given in \cite{HL08a} is flawed.

\section{Flaw in the proof of Corollary 5.1} \label{sec:flaw}

The argument given in \cite{HL08a} is based on the well-known `coupon collector' scenario \cite{Feller57,MCMT00}, where one must complete a collection of $n$ different coupons by acquiring them at random. In the present context, each `coupon' corresponds to a coordinate $i$ of $\vec{p}$, which is `collected' when it is first chosen in \req{eq:P} together with another $j$ such that $(p_i,p_j) \neq (0,0)$. HL carefully show that, if the $Z$ chain has already converged, then after $O(n \ln (n/ \eps))$ circuit steps, the probability that all coordinates have been `hit' in this sense is greater than $1-\eps$. The crux of their argument is however the following statement (p. 284):

\begin{quote}
Once each site of the full chain has been hit, (...) the chain has mixed. This is because, after each site has been hit, the probability distribution over the states is uniform.
\end{quote}

Indeed, if this were true, then standard results, based on the concept of a `strong stationary time' (SST)\footnote{An SST \cite{MCMT00,DiaconisShuf} is an instant $\tau$ when
the distribution $X_{\tau}$ of the chain conditional on a certain event occurring matches the stationary one $\pi$.
 More precisely, $X_{\tau}$ must be obtained
\emph{independently} of $\tau$, and of the initial state $y$ of the chain, ie: $
P_y\{X_{\tau} = x, \tau = t\} = \pi(x)P_y\{\tau = t\}$. 
Under these circumstances a bound on the mixing time can be established (see, e.g., Proposition 6.10 in
\cite{MCMT00}) .} would allow the bound on the `album completion' time $\tau$ to be
converted into one on the $P$ chain's mixing time.
Unfortunately, however, the quoted statement is incorrect: the probability distribution conditioned on all sites being hit is in fact \emph{not} uniform, and an SST-type argument cannot be used.

This is already apparent in \req{eq:P}: note that, conditioned on a site $i$ having just been hit, its value $p_i$ has probability 1/5 of becoming 0 and 4/15 of becoming 1,2 or 3.  In particular, since this is true of the last site to be hit, the overall distribution for $\vec{p}$ conditioned on all sites being hit cannot be uniform.

One can also construct an explicit counterexample. Choose for example $n= 3$ qubits (the simplest nontrivial case) and  initial state $y=(0~0~1)$. Starting from $y$, consider those evolutions such that all three sites are `collected' after two circuit steps. By exhausting all such cases, it is straightforward to check that the conditional probability of reaching each final state is not uniform. For example: the probabilities of obtaining $(1~0~0)$ or $(0~0~1)$ have a ratio 3:2.

\section{Alternative strategy and symmetry analysis}\label{sec:prelims}

While it is conceivable that, with appropriate tweaking, an SST-based argument might still be found for Corollary 5.1, we have been unable to do so.
We propose instead a different strategy, based on reducing the
analysis of the $P$ chain to that of another well-known problem in Markov chain theory: the repeated \emph{random transposition} of $n$ objects.  Note that other kinds of argument may also be possible, for instance via coupling  (A. Harrow, private communication).

 Much is known about the random transposition chain \cite{MCMT00,DiaconisRandomPermut,DiaconisLivroG}; in particular, P. Diaconis and
collaborators have shown that it converges to within $\eps$ of a random permutation after
$\Theta(n \ln(n/\eps))$ steps.\footnote{In fact, much sharper statements can be made
\cite{MCMT00,DiaconisRandomPermut,DiaconisLivroG}, but these are not necessary here.} In order see how this result applies to the problem at hand, let us define the set of states sharing the same Hamming weight $H$:
\beq \label{eq:orbitZ}
G_H : = \left\{ \vec{p} \left| H(\vec{p}) = H\right. \right\}.
\eeq
Since the $Z$ chain mixes after  $\Theta(n \ln(n/\eps))$ circuit steps, then at that point the total probability for each $G_H$ is approximately correct. However, the probability distributions within each set may still be uneven, and so it is not yet possible to ensure that the full $P$ chain has mixed to its uniform stationary state.

Note now that all elements of  $G_H$ are equivalent up to permutations of their indexes and/or of the values 1,2 or 3 of their nonzero coordinates. One can thus expect that applying a random permutation of these variables will result in the mixing of $P$. Lemmas \ref{lem:Q} and \ref{lem:perms} below show that this is indeed true.

The remaining question is then: how do we ensure that such a permutation is applied? A
simple way is to do it `by hand'. For example, once the $Z$ chain has mixed, we can apply
an efficient permutation-generating algorithm such as the Durstenfeld-Knuth shuffle
\cite{Knuth97}, which requires $O(n)$ transpositions to generate an exactly randomly
distributed permutation of the indexes of $\vec{p}$. In physical terms, each
transposition can be implemented by a SWAP gate on the corresponding qubit pair.
Subsequently, all we need is to apply independent permutations of the values $1,2,3$ on
each site. These can all be done in parallel, by applying a random choice from the set of
Pauli rotations $\{\sigma_i\}_{i=1}^3$ on each qubit (compare e.g. the
$\mathcal{C}_1/\mathcal{P}_1$-twirl in \cite{Dankert09}). The overall number of circuit
steps for the entire algorithm is therefore still $\Theta(n \ln(n/\eps))$. Once this is
done, the remainder of the argument  in \cite{HL08a} implies that an approximate quantum
$2$-design will indeed have been generated.

Of course, following this strategy requires switching mid-way from the `pure' random quantum circuit model described by Harrow and Low to a different algorithm. This is  irrelevant if all that is required is an efficient means of generating a $2$-design. Our interest here, however, is to show that the same result is also achieved within the original random circuit model. Specifically, in section \ref{sec:proof} we will show that, once the $Z$ chain has mixed, \emph{the $P$ chain itself} performs the role of a random transposition chain. Diaconis et al's results then ensure that $P$ mixes in $\Theta(n \ln(n/\eps))$ additional steps, and so the overall number of steps will also be of order  $\Theta(n \ln(n/\eps))$.

Before we formalize these ideas, it is useful to exploit the symmetries of $P$ in order
to reduce its analysis to that of a simpler chain, which we call $Q$. This requires some
elementary results from the application of group representation theory to Markov chains
\cite{DiaconisLivroG,BoydDiaconis05}.

\textbf{Markov Chain Projections: }  Suppose a Markov chain $M$, with state space $\Omega_M$, is invariant under an action of some group $G$, i.e.: $M(g(x), g(y)) = M(x,y), \forall g\in G, \forall x,y \in\Omega_M$. If $G_a, G_b \subseteq \Omega_M$ are orbits induced by the group action, then the rule
\beq \label{eq:reducible}
 N(G_a, G_b) : = \sum_{y \in G_b} M(x, y); \; x \in G_a
 \eeq
defines\footnote{Note that this sum is independent of the choice of $x$.} a new Markov
chain $N$ over the set of all orbits, $\{G_i\} \equiv \Omega_N$. This `projected' chain
can be seen as a coarse-graining of the original one. Every probability distribution
$\mu(x)$ over $\Omega_M$ has a natural projection $\nu_{\mu}(a) = \sum_{x \in G_a}
\mu(x)$ on $\Omega_N$. In particular, if $\mu$ is a stationary distribution for $M$, then
$\nu_{\mu}$ is a stationary distribution for $N$. Also, every eigenfunction $h$ of $N$
can be lifted onto a corresponding eigenfunction $f$ of $M$, with the same eigenvalue,
defined by $f(x) : = h(a),\quad  \forall x \in G_a$ (see e.g. Lemma 12.8 in
\cite{MCMT00}). The converse is, in general, not true, since eigenfunctions of $M$ can
project to zero. Thus the projected chain can have fewer eigenvalues than the original
\cite{MCMT00,BoydDiaconis05}, and simpler dynamics. In particular, if both chains are
ergodic, $N$ mixes at least as fast as $M$.

The $Z$ chain is an example of a projection of $P$. By \req{eq:P}, the transition
probabilities $P(\vec{p}, \vec{p}')$ of the $P$ chain are insensitive to whether the
nonzero coordinates of $\vec{p}$ and $ \vec{p}'$ are equal to 1,2 or 3 (they only
distinguish these values from 0). They are also invariant under permutations of the
indexes of $\vec{p}, \vec{p}'$. The group subsuming both these symmetries is isomorphic
to the wreath product\footnote{This is the semidirect product $S_3^n  \rtimes_{\phi}
S_n$, where $\phi$ is the natural homomorphism of $S_3^n$ induced by elements of $S_n$.}
$S_3\wr S_n$. The corresponding orbits  in $\Omega_P$ are precisely the sets $G_H$, and
the projected chain resulting from \req{eq:reducible} is the $Z$ chain.

\textbf{\emph{Q} chain}:  It is useful to define a less coarsely-grained projection of $P$, which we call $Q$,  with state space $\Omega_Q \equiv \{ 0 , 1 \}^n$ (the vertices of a unit hypercube). Consider the action on $\Omega_P$ by the subgroup $S_3^n \subset S_3\wr S_n$ formed by independent permutations of the values $1,2$ and $3$  of each  coordinate of $\vec{p}$. The resulting set of orbits
is isomorphic to  $\Omega_Q $, under the bijection $\vec{q} \leftrightarrow G_{\vec{q}}=~\left\{  \vec{p}\left| p_i =0 \Leftrightarrow q_i = 0 \right.\right\}$.  By eq. (\ref{eq:reducible}), the corresponding projected chain is
\[
Q(\vec{q}, \vec{q}') = \sum_{ \vec{p}' \in G_{\vec{q}'} } P(\vec{p} , \vec{p}') ;\quad  \forall \vec{p} \in G_{\vec{q}}.%
 \]
with stationary state on $\Omega_Q \backslash \{\vec{0} \}$ given by the projection of $\pi$ in \req{eq:Pstat}:
\beq
\nu_{\pi}(\vec{q}) = \frac{1}{4^n-1} 3^{H(\vec{q})}. \label{estacionarioQ}
\eeq
Like $P$, $Q$ may be written as a convex sum $Q = \frac{1}{n(n-1)} \sum_{(i\neq j)} Q^{(i,j)}$, where each $Q^{(i,j)}(\vec{q}, \vec{q'})$ vanishes except for pairs  $\vec{q}, \vec{q'} $ that differ only at coordinates $i$ and $j$. When restricted to these coordinates,  the matrix $Q^{(i,j)}$ always has the same form, given in table~\ref{tab:Q}.

The reason for defining $Q$ is that, despite being a projection of $P$, the two chains have completely equivalent dynamics - we can therefore restrict ourselves to studying the simpler chain\footnote{A related strategy is used in \cite{ODP}}.  As we now show, this happens because the $P$ chain does not distinguish between the elements within each orbit $G_{\vec{q}}$.

\begin{lemma}\label{lem:Q}
The mixing times $t_{mixQ}(\eps)$ and $t_{mixP}(\eps)$ are equal for all $\eps>0$.
\end{lemma}
\emph{Proof:} Since P is a reversible Markov chain, the $t^{th}$ power of its matrix can be expanded as
\[
P^t (\vec{p}, \vec{p}' ) =  \pi(\vec{p}') \sum_{j = 1}^{|\Omega_P|} f_j(\vec{p}) f_j (\vec{p}') \lambda_j^t
\]
where $f_j : \Omega_P \rightarrow  \mathbb{R}^{|\Omega_P|}  \in l^2\left(\pi \right)$ are the eigenfunctions of $P$, with corresponding eigenvalues $\lambda_j$, and which are orthonormal with respect to the stationary measure $\pi$ (see Lemma 12.2 in \cite{MCMT00} ). Similarly,
\[
Q^t (\vec{q}, \vec{q}' ) =  \nu_{\pi}(\vec{q}') \sum_{j = 1}^{|\Omega_Q|} h_j(\vec{q}) h_j (\vec{q}') \alpha_j^t
\]
where $h_j \in l^2\left(\nu_{\pi}\right)$, $\alpha_j$ are the eigenfunctions of $Q$ and corresponding eigenvalues. As previously noted, each $h_j$ can be lifted to a corresponding $f_j$ with same eigenvalue, given by $f_j(\vec{p}) = h_j(\vec{q}), \forall \vec{p} \in G_{\vec{q}}$.

Note now that, by \req{eq:P}
\beq\label{eq:block}
P(\vec{p}_1,\vec{p}'_1) = P(\vec{p}_2,\vec{p}'_2); \quad  \forall \vec{p}_1, \vec{p}_2 \in G_{\vec{q}}, \;\vec{p}'_1, \vec{p}'_2 \in G_{\vec{q}'}.
\eeq
In other words, $P$ can be written as a block-constant matrix, with rank equal to that of $Q$. This implies that the eigenfunctions `lifted' from $h_j$ are the \emph{only} eigenfunctions of $P$ with non-zero eigenvalues. For each $\vec{p} \in G_{\vec{q}}$, we have then
\begin{align}
 \sum_{\vec{p}' \in \Omega_P} \left| P^t (\vec{p}, \vec{p}' )- \pi(\vec{p}') \right| = & \sum_{\vec{p}' \in \Omega_P} \pi(\vec{p}')\left| \sum_{j = 1}^{|\Omega_P|} f_j(\vec{p}) f_j (\vec{p}') \lambda_j^t -1\right| \nonumber \\
= & \sum_{\vec{q}' \in \Omega_Q} \sum_{\vec{p}' \in G_{\vec{q}'}} 3^{-H(\vec{q}')}\nu_{\pi}(\vec{q}')\left| \sum_{j = 1} ^{|\Omega_Q|} h_j (\vec{q}) h_j (\vec{q}') \lambda_j^t - 1\right| \nonumber \\
= &  \sum_{\vec{q}' \in \Omega_Q}  \left| Q^t (\vec{q}, \vec{q}' )- \nu_{\pi}(\vec{q}') \right|
\end{align}
since  there are $3^{H(\vec{q}')}$ elements in $G_{\vec{q}'}$. Finally, since the orbits $G_{\vec{q}}$ for $ \vec{q} \neq \vec{0}$ partition $\Omega_P \backslash{\{\vec{0}\}}$, we obtain the desired result by substituting in  \req{eq:tmix} $\square$

\smallskip

Note that the $Z$ chain is also a projection of the $Q$ chain under the natural action of $S_n$ on $\Omega_Q $, with orbits  $G_H$. Thus every probability distribution  $\nu(\vec{q})$  over $\Omega_Q\backslash\{ \vec{0}\}$ has a projection $\zeta_{\nu}(H)= \sum_{\vec{q} \in G_H} \nu(\vec{q})$ over $\Omega_Z\backslash\{ 0\}$.

\bigskip

We are now ready to formalize the intuitive argument given at the beginning of this
section. Given a permutation $\sigma \in S_n$, let $A_{\sigma}$ be its natural
representation as a Markov matrix acting on the space of probability distributions over
$\Omega_Q$: \footnote{ Here, as is usual in the Markov chain literature \cite{MCMT00},
$\nu$ is a row vector and the Markov matrix $A_{\sigma}$ acts on the left.} \beq [\nu
A_{\sigma}] (\vec{q}) : =  \nu (\sigma(\vec{q}) ) \eeq where
$\left[\sigma(\vec{q})\right]_i  = \vec{q}_{\sigma^{-1} (i)}$ is the natural action of
$\sigma$ on $\Omega_Q$. We can extend this representation to any probability distribution
over $S_n$ by taking convex combinations of the $A_{\sigma}$.  In particular, the uniform
distribution is represented by the Markov matrix $S =\frac{1}{n!} \sum_{\sigma \in S_n}
A_{\sigma}$.

The following lemma shows that applying this random permutation to any distribution $\nu$ over $\Omega_Q$ brings it as close to the stationary state of $Q$ as its projection $\zeta_{\nu}$ is to the stationary state of $Z$.

\begin{lemma} \label{lem:perms}
$ \TVnorm{ \nu S - \nu_{\pi}} = \TVnorm{\zeta_{\nu} - \zeta_{\pi}}$
\end{lemma}

\emph{Proof:} By definition the orbits $G_H$ are invariant under permutations, so
\[
\sum_{\vec{q} \in G_H} [\nu S] (\vec{q}) = \frac{1}{n!} \sum_{\sigma \in S_n} \sum_{\vec{q} \in G_H} [\nu A_{\sigma}] (\vec{q})  =  \sum_{\vec{q} \in G_H} \nu (\vec{q}) = \zeta_{\nu} (H)
\]
Also, since $S A_{\sigma} = S, \forall \sigma$, then $ \nu S$ is a constant function on $G_H$:
\beq \label{eq:unifPerm}
   \nu S (\vec{q}) = \frac{\zeta_{\nu} (H)}{|G_H|}, \forall \vec{q} \in G_H.
\eeq
By \req{estacionarioQ}, the same is also true for $\nu_{\pi}$. Thus, using also \req{eq:reducible}:
 \begin{align}
   \TVnorm{ \nu S - \nu_{\pi}} = & \frac{1}{2} \sum_{H \in \Omega_Z} \sum_{\vec{q} \in G_H}  \left| \nu S (\vec{q}) - \nu_{\pi} (\vec{q})\right| \nn \\
   = & \frac{1}{2} \sum_{H \in \Omega_Z}   \left| \sum_{\vec{q} \in G_H} \nu S (\vec{q}) - \sum_{\vec{q} \in G_H} \nu_{\pi} (\vec{q})\right|  = \TVnorm{\zeta_{\nu} - \zeta_{\pi}} \; _\square
 \end{align}

\begin{table}[tb] \centering \begin{minipage}[b]{0.8\linewidth}
   \begin{center}
     \begin{tabular}{| c | c c c c  | } \hline
    &&& \vspace{-.5cm} & \\
    $Q^{(i,j)}$   & 00 & 01 & 10 & 11 \\  \hline
    00           & 1  & 0  & 0  & 0 \\
    01           & 0  & $1/5$   &  $1/5$   &  $3/5$    \\
    10           & 0  &  $1/5$ & $1/5$& $3/5$ \\
    11           & 0  &  $1/5$   &  $1/5$   & $3/5$\\ \hline
  \end{tabular}
  \hspace{0.3cm}
  \begin{tabular}{| c | c c c c | } \hline
    &&& \vspace{-.5cm} & \\
    $M^{(i,j)}$   & 00 & 01 & 10 & 11 \\  \hline
    00 & 1  &     0    &          0       &         0  \\
    &&& \vspace{-.62cm} & \\
    01 & 0  & $1/4$  &             0              & $3/4$ \\
    10 & 0  &                  0              & $1/4$   & $3/4$ \\
    11 & 0  & $1/4$   & $1/4$   &$1/2$ \\ \hline
  \end{tabular}
  \end{center}
\caption{Transition probabilities $Q^{(i,j)} (q_i q_j, q_i' q_j')$ and $M^{(i,j)} (q_i q_j, q_i' q_j')$. On the left column we have the initial values $q_i q_j$ and on the top line the final values $q_i' q_j'$.} \label{tab:Q}
\end{minipage} \end{table}

\section{Proof of Corollary 5.1 in \cite{HL08a}}
\label{sec:proof}

In this section we show how the $Q$ chain itself induces a random permutation of the
indexes of $\vec{q}$, and how this leads to our desired result, Corollary 5.1 of
\cite{HL08a}. Let us begin by introducing the random transposition chain $\mathcal{T}$
studied by Diaconis et al \cite{MCMT00,DiaconisRandomPermut,DiaconisLivroG}. Consider a
set of $n$ different objects occupying $n$ positions, and subject to the following
evolution rule: at each step, two values $1 \leq i,j \leq n$ are selected independently
at random, and the objects at these positions are swapped. If $i=j$, nothing happens.
Formally, this can be seen as a random walk on
%the permutation group
$S_n$, with transition probabilities between permutations $\sigma$ and $\rho$ given by
 \beq \label{eq:transpchain}
  \mathcal{T}(\sigma, \rho) = \tau\left(\rho \sigma^{-1}\right),
 \eeq
where $\tau$ is the probability distribution over $S_n$ defined by
\begin{equation} \label{eq:RTprobs}
\tau(\alpha) =
      \begin{cases}
        1/n, & \; \alpha = I \\
        2/{n^2}, & \; \alpha \text{ is a transposition} \\
        0,\; & \; \text{otherwise}.
      \end{cases}
\end{equation}
This chain is ergodic and converges to the uniform distribution. As we have already mentioned, Diaconis et al. showed that this occurs with mixing time
\beq \label{eq:Diaconis}
t_{mixT}(\eps) = \Theta(n \ln(n/\eps)).
\eeq

Returning now to the components $Q^{(i,j)}$ of the $Q$ chain (see table~\ref{tab:Q}), notice that each can be rewritten as the convex sum
\begin{equation}
  Q^{(i,j)} = \frac{1}{5}T^{(i,j)} + \frac{4}{5} M^{(i,j)} .
\label{DecomposicaoQ}
\end{equation}
where  $T^{(i,j)}$ represents the transposition of coordinates $i$ and $j$  and $M^{(i,j)}$ is still a Markov matrix. Thus, $Q$ can be seen as the combination of two Markov chains
\beq \label{eq:Qdecomp}
 Q = \frac{1}{5}T_p + \frac{4}{5} M,
\eeq where $T_p \equiv \frac{1}{n(n-1)}\sum_{ i \neq j} T^{(i,j)}$ and $M \equiv
\frac{1}{n(n-1)}\sum_{i\neq j}M^{(i,j)}$, respectively.

The $T_p$ chain represents a random transposition of the components of $\vec{q}$.
Though similar to $\mathcal{T}$, it is based on a different representation of the permutation group: here the transpositions $T^{(i,j)}$ act on the state space $\Omega_Q$, and not $S_n$ itself. As a result, $T_p$ is reducible to independent chains on each of the orbits $G_H$. Furthermore, since $T_p$ lacks the identity component present in \req{eq:RTprobs}, an even (resp. odd) number of steps will always lead to an even (resp. odd) permutation of $\vec{q}$. Thus $T_p$ is a non-convergent, periodic chain.

The latter difficulty can be easily removed by rewriting \req{eq:Qdecomp} as \beq
\label{eq:Qdecomp'} Q=\frac{1}{5} T + \frac{4}{5} \tilde{M} \eeq where $T=\frac{1}{n}I +
\frac{n-1}{n}T_p$ is now aperiodic, and  $\tilde{M} = M +\frac{1}{4n} \left[ T_p - I
\right]$, is an ergodic Markov chain on $\Omega_Q \backslash \{\vec{0}\}$ for $n \geq 3$.
\footnote{This is true since it can be shown \cite{TeseIgor} that i) $M$ is ergodic on
$\Omega_Q \backslash \{\vec{0}\}$ and ii) its eigenvalues are lower-bounded by
$-\frac{2}{3} - \frac{1}{3(n - 1)}$. Thus the eigenvalues of $\tilde{M}$ are all $> -1$
for $n\geq 3$.}

Alternatively,  $T$ may also appear in \req{eq:Qdecomp} if we modify the definition of the two-qubit gate ensemble $\mu$, allowing at each step an extra probability $1/n$ of applying the identity gate. In this case, the $P$ chain in \req{eq:P} becomes  $P'=~\frac{1}{n} I +~\frac{n-1}{n}P$. The corresponding modification of $Q$ leads to
\beq \label{eq:Q'}
Q'=\frac{1}{5}T +\frac{4}{5}\left[ \frac{1 }{ n} I + \frac{n-1}{n}M \right] \equiv \frac{1}{5}T+ \frac{4}{5}M' .
\eeq

The $T$ chain is not ergodic, as it is still reducible into independent chains $T_H$ on each orbit $G_H$. In particular, $T$ does not have a unique stationary state. Nevertheless, it does converge to the random permutation $S$ over $\Omega_Q$, and the mixing time given  in \req{eq:Diaconis} is still valid, in the following generalized sense:

\begin{lemma} \label{lem:tmixGen}
Each initial distribution $\nu$ on $\Omega_Q$ converges under $T$ to its randomized version $\nu S$. In addition, eq. (\ref{eq:Diaconis}) remains valid under the generalized notion
\beq\label{eq:tmixGen}
 t_{mixT} (\eps) = \max_{\nu } \left[ \min t \left| \TVnorm{\nu T^t - \nu S} \leq \eps \right.\right].
\eeq
\end{lemma}
\emph{Proof:} This follows from the fact that each $T_H$ is isomorphic to a projection of $\mathcal{T}$ in the sense of \req{eq:reducible}. See the Appendix for details.

\smallskip

Turning now to the $M$ chain, note from its definition that it is symmetric under permutations of the site indexes, and in particular under transpositions. Thus $M$ (or its variants $\tilde{M}$, $M'$) commutes with $T$.

This property gives us an intuitive picture of how the $Q$ chain behaves. According to \req{eq:Qdecomp'}, each step of $Q$  can be seen as a random choice between moving according to $T$ or to $\tilde{M}$.  A sequence of $t$ steps will, for large enough $t$, contain roughly $t/5$ steps of  $T$ and $4 t / 5$ steps of $\tilde{M}$. Note that the latter are the only steps where the Hamming weight can change, and thus only they contribute to the convergence of the zero chain $Z$. Moreover, since the $T$ and $\tilde{M}$ chains commute, we can consider that all these $\tilde{M}$ steps happen first. Once $Z$ has converged, all we need is to wait for the subsequent $T$ steps to build up to a random permutation of site indexes. Lemmas \ref{lem:Q} and \ref{lem:perms} then ensure that the full $P$ chain will have converged.

Let us now formalize this argument

\smallskip

\textbf{Proof of Corollary 5.1 in \cite{HL08a}}

Let $\gamma >0$, and let $t_0 = t_{mixZ} (\gamma)$, so that the state $\zeta$ of the $Z$ chain after $t_0$ steps satisfies $\TVnorm{\zeta - \zeta_{\pi}}\leq \gamma$. By Lemma \ref{lem:perms}, the corresponding state $\nu$ of the $Q$ chain at that moment lies within the ball
\beq
   B (\gamma) : = \left\{ \nu  \left| \; \TVnorm{\nu S - \nu_{\pi}} \leq \gamma \right. \right\}
\eeq
Define now a mixing time for $Q$ for initial conditions restricted to this ball
\beq\label{eq:tmixBall}
  t_{mixQ} (\eps, \gamma) : =  \max_{\nu \in B (\gamma) } \left[ \min t \left| \;\TVnorm{\nu Q^t - \nu_{\pi}}  \leq \eps \right.\right] 
\eeq
In the Appendix, we show  that this time is bounded by the mixing time  of $T$:
\begin{lemma}\label{lem:tmix}
\beq \label{eq:commutingAB2}
   \sqrt{ t_{mixQ}( \eps, \gamma)} <  \frac{5}{2} \left[ \delta + \sqrt{\delta^2 + \frac{4}{5} \; t_{mixT}\left( \eps - \gamma -  e^{- 2\delta^2 } \right)}\right]
\eeq
 for all $\eps >0 $ and all $\gamma\geq 0, \delta > 0$ satisfying
 $  \eps > e^{- 2\delta^2} + \gamma$.
 \end{lemma}

Choosing $\gamma = \eps/2$, $\delta^2 = \frac{1}{2} \ln (4/ \eps)$ gives
\[
t_{mixQ}( \eps, \eps/2 ) <  \frac{25}{4} \left[ \frac{1}{2} \ln (4/\eps)+ \frac{4}{5} \; t_{mixT}\left( \eps/4  \right)\right]
\]
Taking into account \req{eq:Diaconis} and the fact that $\ln (4/\eps) \leq n \ln (4 n /\eps)\;, \forall n \geq 1$, it follows that there exists an integer $K$ such that
 \[
 t_{mixQ}( \eps, \eps/2 ) < K n \ln (4n/ \eps)  
 \]

Thus the mixing time for the entire $Q$ is
\[
 t_{mixQ}( \eps) \leq t_{mixZ} (\eps/2) + t_{mixQ}( \eps, \eps/2 ) = \Theta( n \ln  (n/ \eps) )
\]
where we use the fact (Theorem 5.1 of \cite{HL08a}) that $t_{mixZ} (\gamma)= \Theta(n \ln (n/\gamma))$.  Finally, applying Lemma \ref{lem:Q} proves our desired result $\square$

\acknowledgements{The authors acknowledge the support of Brazilian funding agencies CNPq and FAPERJ. This work is part of the Brazilian National Institute of Science and Technology of Quantum Information (INCT-IQ). We thank Aram Harrow for encouraging comments, and Roberto I. Oliveira for helpful discussions.}

\appendix*

\section{Proofs of Lemmas 3 and 4}

\subsection{ Proof of Lemma \ref{lem:tmixGen}}

Since all elements of $G_H$ are equivalent under permutations, and transpositions generate all permutations, $T_H$ is irreducible; it is also aperiodic due to the identity component in $T$. Thus, $T_H$ is ergodic, and it is easy to see that its stationary state is the uniform distribution over $G_H$. In other words, any initial distribution $\nu_H$ over $G_H$ converges to $\nu_H S$ (see \req{eq:unifPerm}). Since $T =  \bigoplus_H T_H$, the same is true for any initial distribution $\nu$ over $\Omega_Q$.

Let us now link $T_H$ with Diaconis' $\mathcal{T}$ chain. By the orbit-stabilizer
theorem, $G_H$ is isomorphic to the quotient $S_n / N_H$, where $N_H \subset S_n$ is the
stabilizer of some element  $x^0_H \in G_H$. Explicitly, we identify  $x \in G_H
\leftrightarrow g_xN_H$, where $g_x$ is any permutation such that $g_x(x^0_H) = x$. Since
$T_H$ can be described using the probability distribution $\tau$ in \req{eq:RTprobs}, but
with the transpositions acting on $G_H$, it follows (see, e.g. Lemma 3 in section 3F of
\cite{DiaconisLivroG}) that its transition matrix is
\begin{align} \label{eq:tandT}
T_H(x,y) & = \tau(g_yN_Hg_x^{-1}) =  \mathcal{T}(g_x, g_y N_H)
\end{align}
where we have used \req{eq:transpchain}. Note now that $\mathcal{T}$ is invariant under the action of $N_H$ on $S_n$ given by $h(g) = gh$. The set of orbits of this action is precisely $S_n / N_H \cong G_H$. Comparing \req{eq:tandT} and \req{eq:reducible}, it is clear that $T_H$ is (isomorphic to) the projection of $\mathcal{T}$ with respect to this action. Thus, as discussed in section \ref{sec:prelims}, the mixing time for $T_H$ is at most equal to that of the $\mathcal{T}$, in \req{eq:Diaconis}. Finally, the same is true for $T$ since $T = \bigoplus_H T_H$.

\subsection{Proof of Lemma \ref{lem:tmix}}

Let $p = 1/5$. After $t$ steps of $Q$, the $TV$ distance to the stationary state $\nu_{\pi}$ is, from  \req{eq:Q'}:
\begin{align}
  d(t) &:= \TVnorm{\nu Q^t - \nu_{\pi}  } \nn   = \TVnorm{\nu \sum_{i=0}^t \roundbraket{t \atop i} p^i (1-p)^{t-i} \;  {T}^i \; \tilde{M}^{t-i} -\nu_{\pi} } \nn \\
  & \leq \sum_{i=0}^t \roundbraket{t \atop i}  p^i (1-p)^{t-i}\TVnorm{\nu {T}^i  - \nu_{\pi}}
\label{eq:distanciaL=binomio}
\end{align}
In the first equation we have used the fact that $T$ and $\tilde{M}$ commute, and  in the second the triangle inequality, and also the facts that $\nu_{\pi}$ is the stationary state for $\tilde{M}$, and that applying an ergodic Markov matrix can never increase the $TV$ distance to its stationary state.

Let us now split \req{eq:distanciaL=binomio} into two sums $d_1(t), d_2(t)$, containing respectively terms with $i \leq pt- \delta \sqrt t$ and $i > pt- \delta \sqrt t$, where $\delta>0$ is some constant such that $t > (\delta/p)^2$.  In order to bound $d_1(t)$ we can use the fact that $TV$ distances between probability distributions are always $\leq 1$, so
\[
 d_1(t) \leq \sum_{i=0}^{\left\lfloor pt- \delta \sqrt t\right\rfloor } \roundbraket{t \atop i} p^i (1-p)^{t-i}.
\]
This is a sum of terms in the tail of the binomial distribution, which can again be bound, for any $t> (\delta/p)^2$ using e.g. the Hoeffding inequality \cite{Hoeffding}
\[
d_1(t) \leq \exp( - 2\delta^2).
\]
We can also bound $d_2(t)$, as follows: since $\nu \in B(\gamma)$, then for each value of $i$:
\[
\TVnorm{\nu {T}^i  - \nu_{\pi}}  \leq \TVnorm{\nu {T}^i  - \nu S}  + \TVnorm{\nu S - \nu_{\pi}}  \leq \TVnorm{\nu {T}^i  - \nu S} + \gamma.
\]
Furthermore, since $T$ is ergodic on each orbit $G_H$, and the initial state $\nu$ converges to $\nu S$ by Lemma \ref{lem:tmixGen}, then the $TV$ distance with respect to this state is non-increasing at each step of chain \cite{MCMT00}.  Thus, all $TV$ distances in $d_2$ are at most equal to that of the term with the smallest value $ i=\left\lfloor pt- \delta\sqrt t\right\rfloor  +1$:
\begin{align*}
  d_2 (t) \leq & \left[  \sum^t_{i = \left\lfloor pt- \delta \sqrt t\right\rfloor +1} \roundbraket{t \atop i} p^i (1-p)^{t-i} \right] \Biggl[\TVnorm{\nu T^{\left\lfloor pt- \delta \sqrt t\right\rfloor +1 }  - \nu S }+\gamma \Biggr] \\
  \leq &   \TVnorm{\nu  {T}^{\left\lfloor pt- \delta \sqrt t\right\rfloor+1 }  - \nu S} + \gamma
\end{align*}
since the sum is over part of the binomial distribution. Combining both bounds:
\beq
  d(t) \leq  \exp( - 2\delta^2) + \gamma + \TVnorm{\nu  {T}^{\left\lfloor pt- \delta \sqrt t\right\rfloor+1 }  - \nu S}
  \label{distanciaQ=binomio2}
\eeq

Given now any $\eps > 0$, choose $\gamma, \delta \geq 0 $ satisfying $e^{-2\delta^2} + \gamma < \eps$, and choose also $t$ to be the first instant  for which
\beq \label{eq:tdef}
\max_{\nu \in B(\gamma)} \TVnorm{\nu  {T}^{\left\lfloor pt- \delta \sqrt t\right\rfloor+1 }  - \nu S} \leq \eps - e^{-2\delta^2} - \gamma.
 \eeq
(This instant exists, by Lemma \ref{lem:tmixGen}).  Substituting in \req{distanciaQ=binomio2} and using  \req{eq:tmixBall}:
\beq\label{eq:tmixQ}
 t_{mixQ}( \eps, \gamma )\leq t.
 \eeq
 Define now, in analogy to \req{eq:tmixBall},
\beq\label{eq:tmixTBall}
 t_{mixT} (\eps, \gamma) : =  \max_{\nu \in B (\gamma) } \left[ \min t \left| \;\TVnorm{\nu T^t - \nu S}  \leq \eps \right.\right]  \;  \leq \;t_{mixT}(\eps),
\eeq
with the inequality resulting since $t_{mixT}(\eps)$ maximizes over a larger set. Then we can restate \req{eq:tdef} as
\[pt-~\delta \sqrt t < \left\lfloor pt-~\delta\sqrt{t}\right\rfloor+1  = t_{mixT}\left( \eps - \gamma -  e^{- 2\delta^2 } , \gamma \right).
\]
This inequality, which is quadratic in $\sqrt{t}$, can be inverted to give
\[
\sqrt{t}<  \frac{1}{2p} \left[ \delta + \sqrt{\delta^2 + 4 p\; t_{mixT}\left( \eps - \gamma -  e^{- 2\delta^2 }, \gamma \right)}\right].
\]
Using eqs. (\ref{eq:tmixQ}) and (\ref{eq:tmixTBall}) we obtain the relation between the mixing times of $Q$ and $T$:
\[
\sqrt{t_{mixQ}( \eps, \gamma )} <  \frac{1}{2p} \left[ \delta + \sqrt{\delta^2 + 4 p\; t_{mixT}\left( \eps - \gamma -  e^{- 2\delta^2 } \right)}\right] \quad  \square
\]
\smallskip

\bibliographystyle{unsrt} 
\bibliography{BibDiss}

\end{document}